\documentclass[11pt]{article}
\usepackage{amsfonts,epsfig}
\begin{document}
\begin{sloppypar}
\vspace*{0cm}
\begin{center}
{\setlength{\baselineskip}{1.0cm}{ {\Large{\bf PROPAGATORS OF GENERALIZED SCHR\"ODINGER
EQUATIONS RELATED BY HIGHER-ORDER SUPERSYMMETRY\\}} }} \vspace*{1.0cm}
{\large{\sc{Ekaterina Pozdeeva}}$^\dagger$ and {\sc{Axel
Schulze-Halberg}}$^\ddagger$}
\end{center}
\noindent \\
$\dagger$ Skobeltsyn Institute of Nuclear Physics, Lomonosov Moscow
State University, GSP-1, Leninskie Gory, Moscow 119991, Russian
Federation, e-mail: pozdeeva@www-hep.sinp.msu.ru
\noindent \\ \\
$\ddagger$ Department of Mathematics and Actuarial Science, Indiana University Northwest, 3400 Broadway, Gary, IN 46408, USA,
e-mail: xbataxel@gmail.com

\vspace*{1cm}
\begin{abstract}
\noindent
We construct explicit integral relations between propagators of generalized Schr\"odinger equations that are linked by
higher-order supersymmetry. Our results complement and extend the findings obtained in \cite{bagrovprop}
for the conventional Schr\"odinger equation.

\end{abstract}

\section{Introduction}
This note is directly motivated by the recent work \cite{bagrovprop}, where surprisingly simple
expressions for propagators of supersymmetry-related Schr\"odinger equations were constructed.
In general, the quantum-mechanical supersymmetry (SUSY) formalism connects two Schr\"odinger
equations for (usually) different potentials, such that solutions of one equation are mapped onto solutions
of the second equation \cite{cooper}. The mapping used for relating the solutions to each other is
called SUSY- or Darboux transformation. While the Darboux transformation has been around for more
than a hundred years \cite{darboux}, its application to quantum-mechanical problems within the
SUSY context is much younger,
starting with its relation to the Infeld-Hull factorization \cite{infeld}. Ever since then, the
quantum-mechanical SUSY formalism has been extensively used to study solvability and spectral properties of
Schr\"odinger equations, for an overview the reader may consult \cite{cooper} \cite{junker} \cite{ush}.
Besides the fact that Schr\"odinger equations related by the SUSY formalism allow for mapping their
respective solutions onto each other, such equations are connected in more ways. In particular,
their propagators and the traces of their Green's functions are linked  by relatively simple formulas,
as was shown in \cite{bagrovprop} and \cite{sukumar2} \cite{sukumar}, respectively. It turns out that
the link between the Green's functions even persists under linear generalizations of the
Schr\"odinger equation \cite{green4}, such as the effective mass case or minimal coupling to a
magnetic field. Motivated by this result, in the present note we study the question whether the propagator
relation found in \cite{bagrovprop} can be extended to generalized Schr\"odinger equations. While for
first-order SUSY transformations a positve answer has been given \cite{xbatprop}, here we will study
higher-order transformations. Section 2 gives a short review on the necessary preliminaries, while in section 3
we construct our propagator relations for generalized Schr\"odinger equations.

\section{Preliminaries}
In the following we briefly summarize basic facts about generalized Schr\"odinger equations, the
SUSY formalism, propagators and Green's functions.
\paragraph{The generalized Schr\"odinger equation.} We consider the following generalized Sturm-Liouville
problem on the real interval $(a,b)$, equipped with Dirichlet boundary conditions:
\begin{eqnarray}
f(x)~\psi''(x)+f'(x)~\psi'(x)+[E~h(x)-V(x)]~\psi(x) &=& 0, ~~~x \in (a,b) \label{bvp1} \\
\psi(a) ~~=~~ \psi(b) &=& 0. \label{bvp2}
\end{eqnarray}
Here $f,~h,~V$ are smooth, real functions, with $f,h$ positive and bounded in $a$ and $b$. The constant $E$ will
be referred to as energy, and in solutions of (\ref{bvp1}), (\ref{bvp2}) that belong to the discrete spectrum,
$E$ stands for the spectral value. Any solution $\psi$ of (\ref{bvp1}), (\ref{bvp2}) belonging to a value $E$ from the discrete spectrum, is located in the weighted
Hilbert space $L_h^2(a,b)$ with weight function $h$ \cite{duffy}. The lowest value of the discrete spectrum
will be called the ground state and denoted by $E_0$ with corresponding solution $\psi_0$. The interval $(a,b)$ can be unbounded, that is, $a$ or $b$ can represent minus infinity or infinity, respectively (however,
if $a$ and/or $b$ are finite, then we require $f,~h,~V$ to be continuous there). We see that the problem
(\ref{bvp1}), (\ref{bvp2}) can be singular, which means that its spectrum can admit
a continuous part. Equation (\ref{bvp1}) will be referred to as generalized Schr\"odinger equation, since its
special cases are frequently encountered in Quantum Mechanics, such
as the Schr\"odinger equation for effective mass or with a linearly energy-dependent potential. In the
quantum-mechanical context, $E$ denotes the energy associated with a solution $\psi$, and $V$ stands for
the potential.

\paragraph{Generalized SUSY formalism.} We will now summarize basic facts from \cite{xbatalina2}, for
details see the latter reference. The boundary-value problem (\ref{bvp1}), (\ref{bvp2})
can be linked to another problem of the same kind by means of the SUSY transformation method. Consider
\begin{eqnarray}
f(x)~\Psi''(x)+f'(x)~\Psi'(x)+[E~h(x)-U(x)]~\Psi(x) &=& 0, ~~~x \in (a,b) \label{bvp3} \\
\Psi(a) ~~=~~ \Psi(b) &=& 0, \label{bvp4}
\end{eqnarray}
where the same settings imposed for (\ref{bvp1}), (\ref{bvp2}) apply. Clearly, a solution $\Psi=\Psi(x)$ and the
potential $U=U(x)$ are in general different from their respective counterparts $\psi$ and $V$. Now, suppose
that $\psi$ and $u_0,...,u_{n-1}$ are solutions of the boundary-value problem (\ref{bvp1}), (\ref{bvp2}) and of
equation (\ref{bvp1}) at real energies $E$
and $\lambda_j \leq E$, $j=0,...,n-1$, respectively. Define the $n$-th order SUSY transformation of $\psi$ as
\begin{eqnarray}
D^x_{u_0,...,u_{n-1}}~\psi(x) &=& \left[\frac{f(x)}{h(x)}\right]^\frac{n}{2} \frac{W(u_0,...,u_{n-1},\psi)(x)}
{W(u_0,...,u_{n-1})(x)}, \label{ndarboux}
\end{eqnarray}
where $W$ denotes the Wronskian of the functions in its argument and the upper index $x$ of $D$ denotes
the variable which the derivatives in the Wronskians are applied to. The function
$\Psi=D^x_{u_0,...,u_{n-1}}~\psi$ as defined in (\ref{ndarboux}) solves the boundary-value problem
(\ref{bvp3}), (\ref{bvp4}), if the potential $U$ is given in terms of its counterpart $V$ as follows:
\begin{eqnarray}
U &=& V- 2~f~\frac{d}{dx}\left\{\log\left[W(u_0,...,u_{n-1})\right]\right\}+ 2 ~\frac{d}{dx}
\left(\frac{f}{h}\right) \frac{d}{dx}
\left\{\log\left[W(u_0,...,u_{n-1})\right]\right\}+
\nonumber \\
&+&n~h~\bigg[\frac{f~f'}{2~f~h}-\frac{(f')^2}{2~f~h}-\frac{f'~h'}{2~h^2}+\frac{3~f~(h')^2}{2~h^3}-
\frac{f~h''}{h^2} \bigg]+\nonumber \\
&+&\frac{n^2}{2} ~\bigg[ \frac{(f')^2}{2~f}+\frac{f'~h'}{h}-\frac{3~f~(h')^2}{2~h^2}-
f''+\frac{f~h''}{h} \bigg], \label{potex}
\end{eqnarray}
note that for the sake of brevity the arguments were left out. In the special case $n=1$,
the transformation (\ref{ndarboux}) simplifies to
\begin{eqnarray}
D^x_{u_0}~\psi(x) &=& \sqrt{\frac{f(x)}{h(x)}}~\frac{W(u_0,\psi)(x)}{u_0(x)}
~=~\sqrt{\frac{f(x)}{h(x)}}~\left[-\frac{u_0'(x)}{u_0(x)}~\psi(x)+\psi'(x)\right]. \label{darboux}
\end{eqnarray}
It is well-known \cite{arrigo} \cite{xbatalina2} that the $n$-th order transformation (\ref{ndarboux}) can
always be written as an iteration (or chain) of $n$ first-order transformations (\ref{darboux}), that is,
\begin{eqnarray}
D^x_{u_0,...,u_{n-1}} &=& \prod\limits_{j=0}^{n-1} D^x_{v_j}, \label{factor}
\end{eqnarray}
where each function $v_j$ solves the equation (\ref{bvp1}) that is obtained after the $j$-th first-order SUSY
transformation. Note that (\ref{ndarboux}) and (\ref{darboux}) remain valid when multiplied by a constant, which can be used for
normalization. Now, depending on the choice of the auxiliary solution $u$ in (\ref{darboux}), the discrete spectrum of
problem (\ref{bvp3}), (\ref{bvp4}) can be affected in three possible ways: if $\lambda=E_0$ and $u=\psi_0$, then
$E_0$ is removed from the spectrum of (\ref{bvp3}), (\ref{bvp4}). The opposite case, creation of a new spectral
value $\lambda<E_0$, happens if the auxiliary solution $u$ does not fulfill the boundary conditions
(\ref{bvp4}). Finally, the spectra
of both problems (\ref{bvp1}), (\ref{bvp2}) and (\ref{bvp3}), (\ref{bvp4}) are the same, if we pick
$\lambda<E_0$ and an $u$ that fulfills only one of the boundary conditions (\ref{bvp2}).

\paragraph{Propagator and Green's function.} The propagator governs a quantum system's time evolution.
For a stationary Schr\"odinger equation, the propagator $K$ has the defining property
\begin{eqnarray}
\exp(-i~E~t)~\psi(x) &=& \int\limits_{(a,b)} K(x,y,t)~\psi(y)~dy. \label{defprop}
\end{eqnarray}
Suppose problem (\ref{bvp1}), (\ref{bvp2}) admits a complete set of eigenfunctions $(\psi_n)$, $n=0,1,2,...,M \in
\mathbb{N}_0$, where $M$ can stand for infinity, and $(\phi_k)$, $k \in \mathbb{R}$, belonging to the discrete and the
continuous part of the spectrum, respectively. Then the propagator $K$ has the representation
\begin{eqnarray}
K(x,y,t) = h(y)\left[\sum\limits_{n=0}^M \psi_n(x)~\exp(-i~E_n~t)~\psi_n(y) + \int\limits_{\mathbb{R}}
\phi_k(x)~\exp(-i~k^2~t)~\phi_k(y)~dk \right], \label{prop}
\end{eqnarray}
where $E_n$ and $k^2$ stand for the spectral values belonging to the discrete and continuous spectrum,
respectively. The Green's function $G$ of the problem
(\ref{bvp1}), (\ref{bvp2}) has two equivalent representations \cite{duffy}, both of which we will use here.
In order to state the first representation, let $\psi_{0,l}$ and $\psi_{0,r}$ be solutions of equation
(\ref{bvp1}) that fulfill the following unilateral boundary conditions:
\begin{eqnarray}
\psi_{0,l}(a) ~=~ 0 \qquad \psi_{0,r}(b) ~=~ 0. \label{bvpspec}
\end{eqnarray}
The Wronskian $W_{\psi_{0,l},\psi_{0,r}}$ of these funtions is given by
\begin{eqnarray}
W_{\psi_{0,l},\psi_{0,r}}(x) = \frac{c_0}{f(x)}, \label{wron}
\end{eqnarray}
where $c_0$ is a constant that depends on the explicit form of $\psi_{0,l}$ and $\psi_{0,r}$.
Now we can give the first representation of the Green's function $G_0$ for our boundary value
problem (\ref{bvp1}), (\ref{bvp2}):
\begin{eqnarray}
G(x,y) &=& -\frac{1}{c_0}~\bigg[\psi_{0,l}(y)~\psi_{0,r}(x)~\theta(x-y)+\psi_{0,l}(x)~\psi_{0,r}(y)~\theta(y-x)\bigg],
\label{rep1}
\end{eqnarray}
where $c_0$ is the constant from (\ref{wron}) and $\theta$ stands for the Heaviside distribution. The second
representation of the Green's function $G$ can be obtained as follows, provided
problem (\ref{bvp1}), (\ref{bvp2}) admits a complete set of solutions:
\begin{eqnarray}
G(x,y) &=& \sum\limits_{n=0}^M \frac{\psi_n(x)~\psi_n(y)}{E_n-E} +
\int\limits_{\mathbb{R}} \frac{\phi_k(x)~\phi_k(y)}{k^2-E}~dk,
\label{rep2}
\end{eqnarray}
where the notation is the same as in (\ref{prop}). Note that the Green's function is taken at energy $E$.
\paragraph{Propagators related by first-order SUSY.} In order to obtain a relation between the propagators of the two boundary-value problems
(\ref{bvp1}), (\ref{bvp2}) and (\ref{bvp3}), (\ref{bvp4}), we take the propagator $K_1$ of the second problem and
express it through quantities related to the first problem. For the sake of simplicity we assume for now that
the two boundary-value problems have the same discrete spectrum and that both of them admit
a complete set of solutions belonging to a discrete and a continuous part of the spectrum. Furthermore,
we assume that the solutions of problem (\ref{bvp1}), (\ref{bvp2}) are real-valued functions.
This is no restriction, as equation (\ref{bvp1}) involves only real functions. We then find the following
relations \cite{xbatprop} between the propagators $K_1$ and $K_0$ of our boundary-value problems (\ref{bvp1}),
(\ref{bvp2}), and (\ref{bvp3}), (\ref{bvp4}), respectively:
\begin{eqnarray}
K_1(x,y,t) &=& h(y)~D^x_{u_0}~D^y_{u_0} \int\limits_{(a,b)} K_0(x,z,t)~G_0(z,y)~dz \nonumber \\[1ex]
K_1(x,y,t) &=& h(y)~\left\{D^x_{u_0}~D^y_{u_0} \left[\int\limits_{(a,b)} K_0(x,z,t)~G_0(z,y)~dz \right]
+  \phi_{-1}(x)~\exp(-i~\lambda~t)~\phi_{-1}(y) \right\} \nonumber \\[2ex]
K_1(x,y,t) &=& h(y)~D^x_{u_0}~D^y_{u_0} \int\limits_{(a,b)} K_0(x,z,t)~
\lim\limits_{E \rightarrow E_0} \left[G_0(z,y) - \frac{\psi_0(z)~\psi_0(y)}{E_0-E}\right]
~dz. \label{propsim}
\end{eqnarray}
The first of these relations is valid, if both boundary-value problems admit the same discrete spectrum. The
second relation applies, if (\ref{bvp3}), (\ref{bvp4}) admits an additional discrete spectral value
$\lambda$ with corresponding solution $\phi_{-1}$. Finally, the third relation holds, if the initial problem
(\ref{bvp1}), (\ref{bvp2}) has one discrete spectral value more than its transformed counterpart. In this last case,
if we choose the auxiliary function to be the ground state $\psi_0$ of our initial problem, we can simplify
(\ref{propsim}) as follows:
\begin{eqnarray}
K_1(x,y,t) &=&\sqrt{\frac{h(y)}{f(y)}}~\frac{1}{\psi_0(y)}~ D^x_{\psi_0} \int\limits_{(y,b)} K_0(x,z,t)~\psi_0(z)~dz.
\label{prop0}
\end{eqnarray}
It is immediate to see that the above propagator relations simplify to their well-known conventional
forms \cite{bagrovprop}, if we set $f=h=1$.

\section{Propagators related by higher-order SUSY}
From now on we will assume that a higher-order SUSY transformation (\ref{ndarboux}) was applied to
the initial boundary-value problem (\ref{bvp1}), (\ref{bvp2}), giving the associated problem (\ref{bvp3}),
(\ref{bvp4}). We are looking for an explicit relation between the propagators of these two problems.
To this end, we distinguish whether the SUSY transformation adds new values to the discrete spectrum,
removes some from it or whether the spectra of both boundary-value problems stay the same.

\subsection{Creation of spectral values}
Let us first assume that our $N$-th order SUSY transformation creates a $N$ new discrete spectral values in
the corresponding transformed boundary-value problem. We denote these values and their
corresponding solutions by $E_{-n}$ and $\Psi_{-n}$, $n=1,...,N$, respectively. According to (\ref{propsim}), the propagator of problem
(\ref{bvp3}), (\ref{bvp4}) is then given by
\begin{eqnarray}
K_N(x,y,t) &=& h(y)~\Bigg[\sum\limits_{n=0}^M \Psi_n(x)~\exp(-i~E_n~t)~\Psi_n(y) + \int\limits_{\mathbb{R}}
\Phi_k(x)~\exp(-i~k^2~t)~\Phi_k(y)~dk
+ \nonumber \\
&+& \sum\limits_{n=1}^{N} \Psi_{-n}(x)~\exp(-i~E_{-n}~t)~\Psi_{-n}(y)\Bigg]. \label{propn1}
\end{eqnarray}
Next, we take into account that all functions $\Psi_n$, $n=0,...,M$, and $\Phi_k$, $k \in \mathbb{R}$,
have been obtained from solutions $\psi_j$ by means of an $N$-th order SUSY transformation (\ref{ndarboux}),
using the auxiliary solutions $u_j$, $j=0,...,N-1$:
\begin{eqnarray}
K_N(x,y,t) &=& h(y)~ D^x_{u_0,...,u_{N-1}}~D^y_{u_0,...,u_{N-1}}~\Bigg[\sum\limits_{n=0}^M
L^2_{\psi}~\psi_n(x)~\exp(-i~E_n~t)~\psi_n(y) + \nonumber \\
& & \hspace{-2cm}+~\int\limits_{\mathbb{R}} L^2_{\phi}~\phi_k(x)~\exp(-i~k^2~t)~\phi_k(y)~dk \Bigg]
+  h(y)~\sum\limits_{n=1}^{N} \Psi_{-n}(x)~\exp(-i~E_{-n}~t)~\Psi_{-n}(y), \nonumber \\ \label{kn1}
\end{eqnarray}
where normalization constants $L_{\psi}$ and $L_{\phi}$ were introduced. Before we determine these
constants, we make use of the defining property (\ref{defprop}), transforming (\ref{kn1}) into
\begin{eqnarray}
K_N(x,y,t) &=& h(y)~D^x_{u_0,...,u_{N-1}}~D^y_{u_0,...,u_{N-1}}~\Bigg[\sum\limits_{n=0}^M
L^2_{\psi}~\int\limits_{(a,b)} K_0(x,z,t)~\psi_n(z)~dz
~\psi_n(y) + \nonumber \\
& & \hspace{-3cm} +~\int\limits_{\mathbb{R}} L^2_{\phi}~
\int\limits_{(a,b)} K_0(x,z,t)~\phi_k(z)~dz~\phi_k(y)~dk \Bigg]
+   h(y)~\sum\limits_{n=1}^{N} \Psi_{-n}(x)~\exp(-i~E_{-n}~t)~\Psi_{-n}(y), \nonumber \\ \label{kn2}
\end{eqnarray}
Our normalization constants $L_{\psi}$ and $L_{\phi}$ must be chosen as follows \cite{bagrovsusy}
\begin{eqnarray}
L_\psi ~=~ \prod\limits_{p=0}^{N-1} \sqrt{\frac{1}{E-\lambda_p}} \qquad
L_\phi ~=~ \prod\limits_{p=0}^{N-1} \sqrt{\frac{1}{k^2-\lambda_p}}, \nonumber
\end{eqnarray}
substitution of which renders our propagator relation (\ref{kn2}) in the following form:
\begin{eqnarray}
K_N(x,y,t) &=& h(y)~D^x_{u_0,...,u_{N-1}}~D^y_{u_0,...,u_{N-1}}~\Bigg[\sum\limits_{n=0}^M
\prod\limits_{p=0}^{N-1} \frac{1}{E-\lambda_p}~\int\limits_{(a,b)} K_0(x,z,t)~\psi_n(z)~dz
~\psi_n(y) + \nonumber \\
&+& \int\limits_{\mathbb{R}} \prod\limits_{p=0}^{N-1} \frac{1}{k^2-\lambda_p}~
\int\limits_{(a,b)} K_0(x,z,t)~\phi_k(z)~dz~\phi_k(y)~dk \Bigg]
+  \nonumber \\
&+& h(y)~\sum\limits_{n=1}^{N} \Psi_{-n}(x)~\exp(-i~E_{-n}~t)~\Psi_{-n}(y), \label{kn3}
\end{eqnarray}
Next, we rewrite our normalization constants according to the decomposition
\begin{eqnarray}
\prod\limits_{p=0}^{N-1} \frac{1}{E-\lambda_p}&=&\sum\limits_{p=0}^{N-1}
\prod\limits_{q=0 \atop q \neq p}^{N-1} \frac{1}{\lambda_q-\lambda_p}~\frac{1}{E-\lambda_p} \nonumber \\
\prod\limits_{p=0}^{N-1} \frac{1}{k^2-\lambda_p}&=&\sum\limits_{p=0}^{N-1}
\prod\limits_{q=0 \atop q \neq p}^{N-1} \frac{1}{\lambda_q-\lambda_p}~\frac{1}{k^2-\lambda_p}. \nonumber
\end{eqnarray}
We plug this into our propagator relation (\ref{kn3}) and obtain after regrouping terms
\begin{eqnarray}
K_N(x,y,t) &=& \nonumber \\[2ex]
& & \hspace{-2.5cm} =~ h(y)~D^x_{u_0,...,u_{N-1}}~D^y_{u_0,...,u_{N-1}}~\Bigg\{
\sum\limits_{p=0}^{N-1}
\prod\limits_{q=0 \atop q \neq p}^{N-1} \frac{1}{\lambda_q-\lambda_p}~\Bigg[
\int\limits_{(a,b)} K_0(x,z,t)~\sum\limits_{n=0}^M~\frac{\psi_n(z)~\psi_n(y)}{E-\lambda_p}~dz
 + \nonumber \\[1ex]
& & \hspace{-2.5cm} +~\int\limits_{(a,b)} K_0(x,z,t)~\int\limits_{\mathbb{R}}~\frac{\phi_k(z)~\phi_k(y)}{k^2-\lambda_p}~dk~dz \Bigg]
\Bigg\}
+h(y)~\sum\limits_{n=1}^{N} \Psi_{-n}(x)~\exp(-i~E_{-n}~t)~\Psi_{-n}(y). \label{kn4}
\end{eqnarray}
In the final step we make use of the
representation (\ref{rep2}) of our Green's function, which converts expression (\ref{kn4}) to the form
\begin{eqnarray}
K_N(x,y,t) &=& \nonumber \\
& & \hspace{-1cm} =~h(y)~D^x_{u_0,...,u_{N-1}}~D^y_{u_0,...,u_{N-1}}
~\left[ \sum\limits_{p=0}^{N-1}
\left(\prod\limits_{q=0 \atop q \neq p}^{N-1} \frac{1}{\lambda_q-\lambda_p} \right)
\int\limits_{(a,b)} K_0(x,z,t)~G_0(z,y)~dz \right]+ \nonumber \\
& & \hspace{-1cm}+~h(y)~
\sum\limits_{n=1}^N \phi_{-n}(x)~\exp(-i~E_{-n}~t)~\phi_{-n}(y). \label{propfin1}
\end{eqnarray}
Note that the Green's function depends on the index $n$, as it must be taken at energy $\lambda_n$.
Expression (\ref{propfin1}) reduces correctly to the conventional result \cite{bagrovprop}, if $h=1$ is substituted.

\subsection{Annihilation of spectral values}
We will now assume that our SUSY transformation removes $n$ values from the discrete spectrum. Speaking
in terms of the factored transformation, each step will remove the corresponding ground state from the
current system. In other words, after a SUSY transformation of order $n$, the first $n$ discrete spectral
values of the initial boundary-value problem will have been removed from the discrete spectrum. This
particular ordering of the spectral values and solutions to be deleted does not constitute a restriction
\cite{bagrovprop}, but facilitates notation and calculation. Before we turn to the propagator relations,
it is necessary to set up some notation in order to obtain representations for the
auxiliary solutions. In order to do so, we will consider our higher-order SUSY transformation
(\ref{ndarboux}) in its factored form (\ref{factor}), that is, an $n$-th order transformation is seen as the
$n$-fold application of a first-order transformation. Each first-order transformation yields a new
boundary-value problem of the type (\ref{bvp3}), (\ref{bvp4}), such that in total a sequence of $n$
boundary-value problems is generated. We will assume that in each transformation step the lowest
value is deleted from the discrete spectrum, such that after $n$ iterations of our first-order
transformation the lowest $n$ spectral values are removed. The auxiliary solutions used in each iteration of
our SUSY transformation will be named $u_{j,k}$, where the first index denotes the number of the
boundary-value problem associated with $u_{j,k}$, starting from $j=0$. The second index in $u_{j,k}$
stands for the solution number associated with the $k$-th discrete spectral value. In particular,
$u_{n,n}$ is the ground state of the $n$-th boundary-value problem and the functions $u_{n,j}$ for
$j=n+1,...,M$ represent solutions of
the $n$-th boundary-value problem. Finally,
the $u_{n,j}$ for $j=0,...,n-1$ stand for solutions of the equation associated with our boundary-value
problem at energies that do not belong to the discrete spectrum, which implies that they do not
fulfill the boundary conditions (\ref{bvp4}). For the sake of convenience, let us now introduce
abbreviations for the SUSY transformation operators (\ref{ndarboux}), using the notation for our
auxiliary functions. For natural numbers $n$ and $j\leq n-1$ we define
\begin{eqnarray}
L^x_{n,j} &=& D^x_{u_{j,j},u_{j,j+1},...,u_{j,n-1}}. \label{lnj}
\end{eqnarray}
This operator maps solutions of the $j$-th boundary-value problem onto solutions of the $n$-th
boundary-value problem. Therefore, the operator (\ref{lnj}) admits the contraction property
\begin{eqnarray}
L^x_{n,j}~ L^x_{j,k} &=& L^x_{n,k}. \label{contraction}
\end{eqnarray}
In terms of these operators, our auxiliary solutions $u_{n,j}$ for $j=n,...,M$ are built by the following rule:
\begin{eqnarray}
u_{n,j}(x) &=& L^x_{n+1,j}~u_{n-1,j}(x). \label{auxrule}
\end{eqnarray}
The remaining auxiliary solutions $u_{n,j}$ for $j=0,...,n-1$, which do not fulfill the boundary conditions,
are constructed as follows:
\begin{eqnarray}
u_{n,j}(x) &=& L^x_{n,0}~\hat{u}_{0,j}(x) \nonumber \\[1ex]
\hat{u}_{0,j}(x) &=& u_{0,j}(x) \int \frac{1}{f(x)~u_{0,j}^2(x)}~dx. \nonumber
\end{eqnarray}
Note that the functions $u_{0,j}$ and $\hat{u}_{0,j}$ solve the same equation and are linearly independent
\cite{kamke}. Now we are in position to construct closed-form expressions for our auxiliary solutions.
To this end, we take the well-known expressions from the conventional case $f=h=1$, which were obtained
in \cite{bagrovprop}, and map them to the present, generalized context. Let $v_{n,j}$ stand for the
conventional auxiliary functions, where the indices have the same meaning and vary exactly as in our
generalized case described above, then we have from \cite{bagrovprop}
\begin{eqnarray}
v_{n,j}(x) &=& (E_{n-1}-E_j)~(E_{n-2}-E_j)...(E_{j+1}-E_j)~
\frac{W_n(v_{0,0},...,v_{0,n-1})(x)}{W(v_{0,0},...,v_{0,n-1})(x)}, \label{auxc}
\end{eqnarray}
where the modified Wronskian $W_n$ is obtained from $W$ by removing the $(n+1)$-th row and column
from the underlying matrix. We will now rewrite the auxiliary solutions $v_{n,j}$ and their Wronskians in
(\ref{auxc}) in terms of the present auxiliary functions $u_{n,j}$. In order to do so, we make use of the
following results taken from \cite{xbatalina2}:
\begin{eqnarray}
v_{n,j}(x) &=& \left[f(x)~h(x)\right]^\frac{1}{4}~u_{n,j}(x) \nonumber \\
W_n(v_{0,0},...,v_{0,n-1})(x) &=& \left[f(x)\right]^\frac{(n-1)^2}{4}~\left[h(x)\right]^{-\frac{(n-1)~(n-3)}{4}}
~W_n(u_{0,0},...,u_{0,n-1})(x)
\nonumber \\
W(v_{0,0},...,v_{0,n-1})(x) &=& [f(x)]^\frac{n^2}{4}~[h(x)]^{-\frac{n(n-2)}{4}}~W(u_{0,0},...,u_{0,n-1})(x). \nonumber
\end{eqnarray}
We plug these equalities into (\ref{auxc}), which then becomes
\begin{eqnarray}
& & \hspace{-.8cm}\left[f(x)~h(x)\right]^\frac{1}{4}~u_{n,j}(x) ~= \nonumber \\
& & \hspace{.3cm}=~(E_{n-1}-E_j)~(E_{n-2}-E_j)...(E_{j+1}-E_j)~
\frac{1}{h(x)}~\left[\frac{h(x)}{f(x)} \right]^{\frac{n}{2}-\frac{1}{4}}
\frac{W_n(u_{0,0},...,u_{0,n-1})(x)}{W(u_{0,0},...,u_{0,n-1})(x)}. \nonumber
\end{eqnarray}
Using the abbreviation
\begin{eqnarray}
C_{n,j}&=&(E_{n-1}-E_j)~(E_{n-2}-E_j)...(E_{j+1}-E_j), \nonumber
\end{eqnarray}
renders our auxiliary solution $u_{n,j}$ in the following form:
\begin{eqnarray}
u_{n,j}(x) &=& C_{n,j}~
\frac{1}{h(x)}~\left[\frac{h(x)}{f(x)} \right]^{\frac{n}{2}}
\frac{W_n(u_{0,0},...,u_{0,n-1})(x)}{W(u_{0,0},...,u_{0,n-1})(x)}. \label{auxrep}
\end{eqnarray}
We will now show that the propagator $K_n$ of the boundary-value (\ref{bvp3}), (\ref{bvp4}), which is
obtained after an $n$-chain of SUSY transformations, can be given in the followin form:
\begin{eqnarray}
K_n(x,y,t) &=& \left[\frac{h(y)}{f(y)} \right]^\frac{n}{2} (-1)^{n-1} ~L^x_{n,0}~ \sum\limits_{j=0}^{n-1} ~
\frac{W_j(y)}{W(y)}~\int\limits_{(y,b)} K_0(x,q,t)~u_{0,j}(q)~dq, \label{knk0}
\end{eqnarray}
where $K_0$ is the propagator of the initial boundary-value problem (\ref{bvp1}), (\ref{bvp2}).
In order to establish relation (\ref{knk0}), we will use induction, proceeding similarly to how it was done
in the conventional case \cite{bagrovprop}. The first-order case $K_1$ has already been established
\cite{xbatprop} and is given in (\ref{prop0}). Now assume this relation to hold for $K_n$, then our induction
step starts at
\begin{eqnarray}
K_{n+1}(x,y,t) &=&\sqrt{\frac{h(y)}{f(y)}}~\frac{1}{u_{n,n}(y)}~ L^x_{n+1,n}~ \int\limits_{(y,b)} K_n(x,z,t)
~u_{n,n}(z)~dz. \label{kx1}
\end{eqnarray}
Since we assume that (\ref{knk0}) is true for $n$, we can substitute it into (\ref{kx1}). After ordering terms, we
get
\begin{eqnarray}
K_{n+1}(x,y,t) &=& \nonumber \\[1ex]
& & \hspace{-3cm}=~\sqrt{\frac{h(y)}{f(y)}}~\frac{(-1)^{n-1}}{u_{n,n}(y)}~ L^x_{n+1,n}~L^x_{n,0}
 \sum\limits_{j=0}^{n-1} \int\limits_{(y,b)}
 \left(\frac{h(z)}{f(z)} \right)^\frac{n}{2}
\frac{W_j(z)}{W(z)}~u_{n,n}(z) \int\limits_{(z,b)} K_0(x,q,t)~u_{0,j}(q)~dq~dz . \nonumber
\end{eqnarray}
Next, we substitute the ratio of Wronskians $W_j/W$ by means of our representation (\ref{auxc}),
make use of the contraction (\ref{contraction}) and arrive after some simplification at
\begin{eqnarray}
K_{n+1}(x,y,t) &=& \nonumber \\[1ex]
& & \hspace{-2.8cm}=~ \sqrt{\frac{h(y)}{f(y)}}~\frac{(-1)^{n-1}}{u_{n,n}(y)}~L^x_{n+1,0} \sum\limits_{j=0}^{n-1}
\frac{(-1)^j}{C_{n,j}}~\int\limits_{(y,b)}  h(z)~u_{n,n}(z)~u_{n,j}(z) ~\int\limits_{(z,b)}K_0(x,q,t)~u_{0,j}(q)~dq~dz.  \nonumber
\end{eqnarray}
The area of integration forms a triangle $T$ in $z$-$q$-space, which can be seen as the upper half of the
rectangle $[y,b] \times [y,b]$. Therefore, integration over $T$ can be replaced by integration over the
rectangle minus integration over its lower triangle:
\begin{eqnarray}
K_{n+1}(x,y,t) &=& \sqrt{\frac{h(y)}{f(y)}}~\frac{(-1)^{n-1}}{u_{n,n}(y)}~L^x_{n+1,0} \sum\limits_{j=0}^{n-1}
\frac{(-1)^j}{C_{n,j}} \times \nonumber \\
&\times& \left[ ~\int\limits_{(y,b)}  h(z)~u_{n,n}(z)~u_{n,j}(z)~dz ~\int\limits_{(y,b)}K_0(x,q,t)~u_{0,j}(q)~dq-
\right. \nonumber \\
&-& \left.  \int\limits_{(y,b)}K_0(x,q,t)~u_{0,j}(q)~\int\limits_{(q,b)}  h(z)~u_{n,n}(z)~u_{n,j}(z) ~dz~dq \right].
\label{kx2}
\end{eqnarray}
Before we process this expression further, let us evaluate the integral with respect to $z$:
\begin{eqnarray}
\int\limits_{(\xi,b)}  h(z)~u_{n,n}(z)~u_{n,j}(z)~dz &=& \frac{f(\xi)~W(u_{n,n},u_{n,j})(\xi)}{E_j-E_n}-
\frac{f(b)~W(u_{n,n},u_{n,j})(b)}{E_j-E_n}. \label{kx3}
\end{eqnarray}
This can be verified in a straightforward way by diffentiating the right hand side and replacing the
second derivatives by means of our equation (\ref{bvp1}). The second term on the right-hand side of
(\ref{kx3}) is a constant and will cancel out in subsequent calculations, whereas the first term will now
be modified further. First we make use of the relation
\begin{eqnarray}
L^\xi_{n+1,n} u_{n,j}(\xi) &=& \sqrt{\frac{f(\xi)}{h(\xi)}}~\frac{W(u_{n,n},u_{n,j})(\xi)}{u_{n,n}(\xi)}, \nonumber
\end{eqnarray}
replacing the Wronskian in (\ref{kx3}):
\begin{eqnarray}
\int\limits_{(\xi,b)}  h(z)~u_{n,n}(z)~u_{n,j}(z)~dz &=& \frac{\sqrt{f(\xi)~h(\xi)}~u_{n,n}(\xi)~L^\xi_{n+1,n} u_{n,j}(\xi)}{E_j-E_n}-
\frac{f(b)~W(u_{n,n},u_{n,j})(b)}{E_j-E_n} \nonumber \\
&=& \frac{\sqrt{f(\xi)~h(\xi)}~u_{n,n}(\xi)~u_{n+1,j}(\xi)}{E_j-E_n}-
\frac{f(b)~W(u_{n,n},u_{n,j})(b)}{E_j-E_n}. \label{kx4}
\end{eqnarray}
We now replace the function $u_{n+1,j}$ by its representation (\ref{auxrep}) for $n+1$ and simplify the result:
\begin{eqnarray}
& & \hspace{-.5cm} \int\limits_{(\xi,b)}  h(z)~u_{n,n}(z)~u_{n,j}(z)~dz ~= \nonumber \\
& & \hspace{1cm} =~-C_{n,j}~u_{n,n}(\xi) \left[\frac{h(\xi)}{f(\xi)} \right]^{\frac{n}{2}} \frac{W_{j}(u_{0,0},...,u_{0,n})(\xi)}{W(u_{0,0},...,u_{0,n})(\xi)}-
\frac{f(b)~W(u_{n,n},u_{n,j})(b)}{E_j-E_n}. \label{kx5}
\end{eqnarray}
We can replace the integral (\ref{kx5}) in our propagator relation (\ref{kx2}), note that the integral appears twice
there. After some simplification and regrouping terms, we arrive at the following expression:
\begin{eqnarray}
K_{n+1}(x,y,t) &=& \left[\frac{h(y)}{f(y)}\right]^{\frac{n+1}{2}}(-1)^{n}~L^x_{n+1,0} \sum\limits_{j=0}^{n-1}
(-1)^j~\frac{W_{j}(y)}{W(y)}~\int\limits_{(y,b)}K_0(x,q,t)~u_{0,j}(q)~dq+ \nonumber \\
& & \hspace{-2.5cm}+~\sqrt{\frac{h(y)}{f(y)}}~\frac{(-1)^{n-1}}{u_{n,n}(y)}~L^x_{n+1,0}~\int\limits_{(y,b)}
K_0(x,q,t)~\left[\frac{h(y)}{f(y)}\right]^{\frac{n}{2}} \frac{u_{n,n}(q)}{W(q)}
~\sum\limits_{j=0}^{n-1} (-1)^j~u_{0,j}(q)~W_j(q)~dq. \nonumber \\ \label{kx6}
\end{eqnarray}
We will now simplify the second line of this expression for $K_{n+1}$. The last sum can be seen as
an application of Laplace's theorem:
\begin{eqnarray}
\sum\limits_{j=0}^{n-1} (-1)^j~u_{0,j}(q)~W_j(q) &=& (-1)^{n+1}~u_{0,n}(q)~W(u_{0,0},...,u_{0,n-1})(q),
\label{endsum}
\end{eqnarray}
that is, the determinant $W(u_{0,0},...,u_{0,n-1})$ is expanded with respect to its first row. Next, we combine
(\ref{endsum}) with its two factors in front, expressing $u_{n,n}$ by means of (\ref{auxrule}) and
(\ref{ndarboux}):
\begin{eqnarray}
u_{n,n}(q) &=& \left[\frac{f(q)}{h(q)}\right]^\frac{n}{2} \frac{W(u_{0,0},...,u_{0,n})(q)}
{W(u_{0,0},...,u_{0,n-1})(q)}, \label{groundrep}
\end{eqnarray}
which leads to the following result:
\begin{eqnarray}
\left[\frac{h(y)}{f(y)}\right]^{\frac{n}{2}} \frac{u_{n,n}(q)}{W(q)}
~\sum\limits_{j=0}^{n-1} (-1)^j~u_{0,j}(q)~W_j(q) &=& (-1)^{n+1}~u_{0,n}(q). \nonumber
\end{eqnarray}
After substitution of this expression into (\ref{kx6}), we get the following form of our propagator
relation:
\begin{eqnarray}
K_{n+1}(x,y,t) &=& \left[\frac{h(y)}{f(y)}\right]^{\frac{n+1}{2}}(-1)^{n}~L^x_{n+1,0} \sum\limits_{j=0}^{n-1}
(-1)^j~\frac{W_{j}(y)}{W(y)} \int\limits_{(y,b)}K_0(x,q,t)~u_{0,j}(q)~dq+ \nonumber \\
& & \hspace{-2.5cm}+~\sqrt{\frac{h(y)}{f(y)}}~\frac{1}{u_{n,n}(y)}~L^x_{n+1,0} \int\limits_{(y,b)}
K_0(x,q,t)~u_{0,n}(q)~dq. \nonumber
\end{eqnarray}
Using once more our representation (\ref{groundrep}) for $u_{n,n}$, we obtain
\begin{eqnarray}
K_{n+1}(x,y,t) &=& \left[\frac{h(y)}{f(y)}\right]^{\frac{n+1}{2}}(-1)^{n}~L^x_{n+1,0} \sum\limits_{j=0}^{n-1}
(-1)^j~\frac{W_{j}(y)}{W(y)}~\int\limits_{(y,b)}K_0(x,q,t)~u_{0,j}(q)~dq+ \nonumber \\
& & \hspace{-2.5cm}+\left[\frac{h(y)}{f(y)}\right]^\frac{n+1}{2} \frac{W(u_{0,0},...,u_{0,n-1})(y)}
{W(u_{0,0},...,u_{0,n})(y)}~L^x_{n+1,0}~\int\limits_{(y,b)}
K_0(x,q,t)~u_{0,n}(q)~dq. \nonumber
\end{eqnarray}
The second line of the right-hand side turns out to be the $n$-th term of the sum in the first line, such that
we get the following final form of our propagator relation:
\begin{eqnarray}
K_{n+1}(x,y,t) &=&  \left[\frac{h(y)}{f(y)}\right]^{\frac{n+1}{2}}(-1)^{n}~L^x_{n+1,0} \sum\limits_{j=0}^{n}
(-1)^j~\frac{W_{j}(y)}{W(y)}~\int\limits_{(y,b)}K_0(x,q,t)~u_{0,j}(q)~dq. \nonumber
\end{eqnarray}
This expression coincides with (\ref{kx1}) if $n$ is set to $n+1$, and the induction is complete.

\subsection{Isospectrality}
We will now consider the remaining case of a SUSY transformation that renders the discrete
spectrum of the initial boundary-value problem (\ref{bvp1}), (\ref{bvp2}) and its final counterpart
(\ref{bvp3}), (\ref{bvp4}) the same. In particular, we assume that in
each transformation step of the chain the discrete spectrum is preserved. Consequently, the auxiliary
functions $u_0,...,u_{n-1}$ used in the SUSY transformation (\ref{ndarboux}) fulfill only one of the
boundary conditions (\ref{bvp2}). Adopting the settings from \cite{bagrovprop}, from now on we will
require our boundary-value problem to be defined on the whole real line, that is, the quantities
$a$, $b$ in (\ref{bvp2}) stand for negative infinity and infinity, respectively. For the sake of simplicity let
us first assume that our auxiliary solutions $u_0,...,u_{n-1}$ only fulfill the first boundary condition, that is,
\begin{eqnarray}
\lim\limits_{x \rightarrow -\infty} u_j(x) &=& 0. \label{aux1}
\end{eqnarray}
The general case of some $u_j$ fulfilling the first boundary condition, and some fulfilling the second
boundary condition will arise easily once we have established our findings for the setting (\ref{aux1}).
More precisely, we will now prove that the following relation between the propagators $K_0$ and
$K_n$ of our boundary-value problems (\ref{bvp1}), (\ref{bvp2}) and (\ref{bvp3}), (\ref{bvp4}) holds:
\begin{eqnarray}
K_n(x,y,t) &=& (-1)^n \left[\frac{h(y)}{f(y)}\right]^\frac{n}{2} L^x_{n,0} ~\sum\limits_{j=0}^{n-1}
(-1)^j~\frac{W_j(y)}{W(y)}~\int\limits_{-\infty}^y K_0(x,z,t)~u_j(z)~dz. \label{4kn}
\end{eqnarray}
The proof of this propagator relation will follow the same steps that were taken in \cite{bagrovprop}.
We must show that $K_n$ satisfies the generalized time-dependent Schr\"odinger equation associated to
the transformed boundary-value problem (\ref{bvp3}), (\ref{bvp4}), and that for $t=0$ the propagator
(\ref{4kn}) becomes a delta. In order to prove this latter statement, let us evaluate (\ref{4kn}) for $t=0$.
Introducing the Heaviside distribution $\theta$, we have
\begin{eqnarray}
K_n(x,y,0) &=& \nonumber \\
& & \hspace{-1cm} =~(-1)^n~\left[\frac{h(y)}{f(y)}\right]^\frac{n}{2}~L_{n,0}^x~\sum\limits_{j=0}^{n-1}
~(-1)^j~\frac{W_n(y)}{W(y)}~\int\limits_{-\infty}^y K_0(x,z,0)~u_j(z)~dz
\nonumber \\
& & \hspace{-1cm} =~ (-1)^n~\left[\frac{h(y)}{f(y)}\right]^\frac{n}{2}~L_{n,0}^x~ \sum\limits_{j=0}^{n-1}~
(-1)^j~\frac{W_n(y)}{W(y)}~ \theta(y-x)~u_j(x)
\nonumber \\
& & \hspace{-1cm} =~ (-1)^n~\left[\frac{h(y)}{f(y)}\right]^\frac{n}{2}~\sum\limits_{j=0}^{n-1}~
(-1)^j~\frac{W_n(y)}{W(y)}~ L^x_{n,0} \left(\theta(y-x)~u_j(x) \right)
\nonumber \\
& & \hspace{-1cm} =~ (-1)^n~\left[\frac{h(y)}{f(y)}\right]^\frac{n}{2}~\sum\limits_{j=0}^{n-1}~
(-1)^j~\frac{W_n(y)}{W(y)}~ \left(\frac{f(x)}{h(x)}\right)^\frac{n}{2}
\frac{W(u_0,...,u_{j-1},\theta(y-x)~u_j)(x)}{W(u_0,...,u_{j-1})(x)}.
\nonumber \\ \label{4kn2}
\end{eqnarray}
Let us now rewrite the Wronskian that involves the Heaviside distribution. Note that the following
argument is the same that was used in \cite{bagrovprop}, we include it here for the sake of
completeness. In order to rewrite the last Wronskian in (\ref{4kn2}), we first need to consider the
derivatives:
\begin{eqnarray}
\frac{\partial^m}{\partial x^m} ~\theta(y-x)~u_j(x) &=&
\sum\limits_{k=0}^{m-1} c_{km}~\left[\frac{\partial^{m-k}}{\partial x^{m-k}}~ \theta(y-x) \right]
~u^{(k)}_j(x)+\theta(y-x)~u^{(m)}_j(x), \label{der}	
\end{eqnarray}
where the $c_{km}$ denote constants. If we use this expression to replace the
derivatives in the last Wronskian of (\ref{4kn2}) and
convert the sum in (\ref{der}) into a sum of two Wronskians, then the last term vanishes, as
$W(u_0,...,u_{n-1},u_j)=0$ for $j=0,...,n-1$. We then arrive at
\begin{eqnarray}
W(u_0,...,u_{j-1},\theta(y-x)~u_j)(x) &=& \nonumber \\[2ex]
& & \hspace{-4cm}=~ \mbox{det}
\left(\begin{array}{cccl}
u_0(x) & \cdots & u_{n-1}(x) & 0 \\
u'_0(x) & \cdots & u'_{n-1}(x) & -\delta(x-y)~u_j(x) \\
\vdots & \ddots & \vdots & \vdots \\
u^{(n)}_0(x) & \cdots & u^{(n)}_{n-1}(x) &
\sum\limits_{k=0}^{n-1} c_{kn} \left(\frac{\partial^{n-k}}
{\partial x^{n-k}}~ \theta(y-x) \right) u^{(k)}_j(x)
\end{array}
\right). \label{4kn3}
\end{eqnarray}
Let us point out that each element of the last column is the sum given by the entry in the lower right
corner, evaluated at $n=0,1,2,...$. Evaluation of the determinant in (\ref{4kn3}) with respect to the
last column gives
\begin{eqnarray}
W(u_0,...,u_{j-1},\theta(y-x) u_j)(x) &=& \nonumber \\
& & \hspace{-3cm}=~ (-1)^n~\sum\limits_{m=0}^n (-1)^m~W_{nm}(x)
\sum\limits_{k=0}^{m-1} c_{km} \left[\frac{\partial^{m-k}}
{\partial x^{m-k}}~ \theta(y-x) \right] u^{(k)}_j(x),  \label{4kn4}
\end{eqnarray}
where as usual the $W_{nm}$ denote determinants of the minor matrices obtained by deleting the $n$-th row and
the $m$-th column. We will now substitute this result into our propagator (\ref{4kn2}) and show that it gives a delta, which is
equivalent to
\begin{eqnarray}
\int\limits_\mathbb{R} K_n(x,y,0)~\phi(x)~dx &=& \phi(y), \label{intprop}
\end{eqnarray}
for all admissible test functions $\phi$, recall that these functions are smooth and vanish at the infinities.
After substitution of (\ref{4kn4}) into (\ref{4kn2}), we obtain
\begin{eqnarray}
\int\limits_\mathbb{R} K_n(x,y,0)~\phi(x)~dx &=&  \left[\frac{h(y)}{f(y)}\right]^\frac{n}{2}
\sum\limits_{j=0}^{n-1} \sum\limits_{m=1}^{n}\sum\limits_{k=0}^{m-1} (-1)^{j+m}~c_{mk}~
\frac{W_j(y)}{W(y)} \times \nonumber \\
&\times& \int\limits_{\mathbb{R}} \left[\frac{\partial^{m-k}}
{\partial x^{m-k}}~ \theta(y-x) \right] \left[\frac{h(y)}{f(y)}\right]^\frac{n}{2}
\frac{W_{nm}(x)~f(x)~u_j^{(k)}(x)}{W(x)}~dx. \label{4kn5}
\end{eqnarray}
The derivatives applied to the Heaviside function can be removed by using $\theta'=\delta$ in the
distributional sense. Taking into account $\delta(y-x)=\delta(x-y)$ and applying the definition of
the delta function's derivative, we obtain from (\ref{4kn5})
\begin{eqnarray}
\int\limits_\mathbb{R} K_n(x,y,0)~\phi(x)~dx &=& \left[\frac{h(y)}{f(y)}\right]^\frac{n}{2}
\sum\limits_{j=0}^{n-1} \sum\limits_{m=1}^{n}\sum\limits_{k=0}^{m-1} (-1)^{j+m}~c_{mk}~
\frac{W_j(y)}{W(y)} \times \nonumber \\
&\times& \int\limits_{\mathbb{R}} \left[\frac{\partial^{m-k-1}}
{\partial x^{m-k-1}}~ \delta(y-x) \right] \left[\frac{h(x)}{f(x)}\right]^\frac{n}{2}
\frac{W_{nm}(x)~f(x)~u_j^{(k)}(x)}{W(x)}~dx \nonumber \\
&=& \left[\frac{h(y)}{f(y)}\right]^\frac{n}{2}
\sum\limits_{j=0}^{n-1} \sum\limits_{m=1}^{n}\sum\limits_{k=0}^{m-1} (-1)^{j-k}~c_{mk}~
\frac{W_j(y)}{W(y)} \times \nonumber \\
&\times& \int\limits_{\mathbb{R}} \frac{\partial^{m-k-1}}
{\partial y^{m-k-1}} \left\{\left[\frac{h(y)}{f(y)}\right]^\frac{n}{2}
\frac{W_{nm}(y)~f(y)~u_j^{(k)}(y)}{W(y)}\right\} dy. \nonumber \\ \label{4kn6}
\end{eqnarray}
We now apply the Leinniz rule to the last term on the right-hand side and make use of the rule
\begin{eqnarray}
\sum\limits_{j=0}^{n-1} (-1)^j~\frac{W_n(y)}{W(y)}~u_j^{(s)}(y) &=& \delta_{s,n-1}, \nonumber
\end{eqnarray}
which turns (\ref{4kn6}) into
\begin{eqnarray}
\int\limits_\mathbb{R} K_n(x,y,0)~\phi(x)~dx &=&
(-1)^n~\frac{W_{nn}(y)~f(y)}{W(y)}~\sum\limits_{k=0}^{n-1} c_{nk}~(-1)^k \label{4kn7} \\
&=& f(y), \nonumber
\end{eqnarray}
because $W_{nn}=W$ and the sum in (\ref{4kn7}) equals $(-1)^n$, see e.g. \cite{bagrovprop}.
Hence, we have shown that $K_n(x,y,0)=\delta(x-y)$. It remains to prove that the propagator
$K_n$ solves the time-dependent Schr\"odinger equation associated with our generalized
boundary-value problem (\ref{bvp3}), (\ref{bvp4}), with respect to both spatial variables $x$ and $y$.
This is obvious in the first case, as $K_n$ is obtained from $K_0$ by application of $L^x_{n,0}$, which
maps solutions of our initial boundary-value problem (\ref{bvp1}), (\ref{bvp2}) onto its transformed
counterpart (\ref{bvp3}), (\ref{bvp4}). Since $K_0$ solves the initial problem, it follows that $K_n$ must
solve the transformed problem. Regarding the second variable $y$ we must substitute the explicit form of
$K_n$, as given in (\ref{4kn}) into the time-dependent Schr\"odinger equation and show that it it fulfilled. This
short calculation follows exactly the same steps as in \cite{bagrovprop}, such that we omit to show it here.
Let us now consider the case where the auxiliary solutions in our SUSY transformation fulfill the
second boundary condition in (\ref{bvp2}), that is,
\begin{eqnarray}
\lim\limits_{x \rightarrow \infty} u_j(x) &=& 0. \nonumber
\end{eqnarray}
In this case, one uses the same argumentation as given above
and arrives as the propagator relation
\begin{eqnarray}
K_n(x,y,t) &=& (-1)^{n-1} \left[\frac{h(y)}{f(y)}\right]^\frac{n}{2} L^x_{n,0} ~\sum\limits_{j=0}^{n-1}
(-1)^j~\frac{W_j(y)}{W(y)}~\int\limits_{y}^\infty K_0(x,z,t)~u_j(z)~dz. \nonumber
\end{eqnarray}
Finally, if the first $M$ auxiliary solutions fulfill the first boundary condition, and the remaining $N-M$
auxiliary solutions satisfy the second boundary condition, then our propagator relation will read
\begin{eqnarray}
K_n(x,y,t) &=& (-1)^{n} \left[\frac{h(y)}{f(y)}\right]^\frac{n}{2} L^x_{n,0} ~\sum\limits_{j=0}^{M-1}
(-1)^j~\frac{W_j(y)}{W(y)}~\int\limits_{-\infty}^y K_0(x,z,t)~u_j(z)~dz+\nonumber \\
&+& (-1)^{n-1} \left[\frac{h(y)}{f(y)}\right]^\frac{n}{2} L^x_{n,0} ~\sum\limits_{j=M}^{n-1}
(-1)^j~\frac{W_j(y)}{W(y)}~\int\limits_{y}^\infty K_0(x,z,t)~u_j(z)~dz. \nonumber
\end{eqnarray}
As before, it is straightforward to see that our propagator relations reduce correctly to the conventional ones
constructed in \cite{bagrovprop}, if we set $f=h=1$.

\section{Concluding remarks}
We have shown that the relations between propagators of SUSY-linked Schr\"odinger equations extend to
the linearly generalized case. In particular, all possible SUSY scenarios (creation, annihilation of spectral values
and isospectrality) have been verified and found to match corresponding findings in \cite{bagrovprop}.

\end{sloppypar}

\end{document}